\documentclass[a4paper,12pt]{article}

\usepackage[T2A]{fontenc}
\usepackage[cp1251]{inputenc}
\usepackage{amsmath}

\usepackage{amsfonts}
\usepackage{epsfig}
\usepackage{booktabs}
\usepackage{floatflt}

\numberwithin{equation}{section}

\newcommand{\la}{\label}
\newcommand{\nn}{\nonumber}
\newcommand{\bea}{\begin{array}}
\newcommand{\eea}{\end{array}}

\newcommand{\nnr}{\nonumber\\}

\newcommand{\ex}{{\rm e}}
\newcommand{\A}{\alpha}

\newcommand{\G}{\Gamma}

\newcommand{\CP}{{\mathbb C \mathbb P}}
\newcommand{\beq}{\begin{equation}}
\newcommand{\eeq}{\end{equation}}
\newcommand{\fr}{\frac}
\newcommand{\beqn}{\begin{eqnarray}}
\newcommand{\eeqn}{\end{eqnarray}}
\newcommand{\pt}{\partial}

\newcommand{\eps}{\epsilon}
\newcommand{\vphi}{\varphi}

\title{Tuning the Circular Strings}
\author{}
\date{}
\begin{document}
\begin{titlepage}

\hfill\parbox{40mm}
{\begin{flushleft}  ITEP-TH-77/09
\end{flushleft}}

\vspace{30mm}

\begin{center}
{\large \bf On the Fermionic Frequencies of Circular Strings}

\vspace{17mm}

\textrm{Victor~Mikhaylov}
\vspace{8mm}

\textit{Institute for Theoretical and Experimental Physics,\\
B.~Cheremushkinskaya ul. 25, 117259 Moscow, Russia;\\
Moscow Institute of Physics and Technology,\\
Institutsky per. 9, 141 700 Dolgoprudny, Russia}\\
\texttt{victor.mikhaylov AT gmail.com}

\vspace{3.5cm}

{\bf Abstract}\end{center}
We revisit the semiclassical computation of the fluctuation spectrum around different circular string solutions in $AdS_5\times S^5$ and $AdS_4\times\CP^3$, starting from the Green-Schwarz action. It has been known that the results for these frequencies obtained from the algebraic curve and from the worldsheet computations sometimes do not agree. In particular, different methods give different results for the half-integer shifts in the mode numbers of the frequencies. We find that these discrepancies can be removed if one carefully takes into account the transition matrices in the spin bundle over the target space.

\end{titlepage}

\section{Introduction}
In this note we revisit the computation of the semiclassical spectrum of fluctuations for circular strings. The semiclassical frequencies have been computed for various string configurations both in $AdS_5\times S^5$ and in $AdS_4\times\CP^3$, by direct expansion of the worldsheet Green-Schwarz action (see {\it e.g.} \cite{wssu2,wssl2,TheThree,McLRTs,Chuvaki}), and using the algebraic curve technique (see {\it e.g.} \cite{GV07,Gromov:2008ec,GM,Chuvaki}). In general the results of the two methods agree, but still there are some discrepancies.

A typical frequency for a fluctuation of polarization $(i,j)$ and mode number $n$ has the form
\beq
\omega_n^{ij}=\omega^{ij}_0+\Omega^{ij}(n-n^{ij}_0)\,,
\eeq
where $\omega_0$ is some constant shift, $n_0$ is a shift of the mode number (in general, half-integer), and $\Omega^{ij}(n)$ is some function. Worldsheet (WS) and algebraic curve (AC) approaches give the same function $\Omega^{ij}(n)$, but the constants $\omega_0$ and $n_0$ sometimes do not agree. The {\it integer} part of $n_0$ is a matter of labeling the frequencies, or a matter of choosing the cutoffs in the sum over $n$ that gives the one-loop correction to the energy.
For the discussion of regularization and prescriptions for labeling the frequencies see \cite{Gromov:2007cd, GM, McLRTs, Chuvaki}. 

In this paper we discuss the {\it half-integer} shifts $n^{ij}_0$, which cannot be absorbed into relabeling. One can note that there are contradictions in the literature concerning the half-integer shifts for fermionic modes of circular strings (see App. E of \cite{GV07}). The results of worldsheet and algebraic curve computations can be found in tab.\ref{tab1} and tab.\ref{tab2}. For the $sl(2)$ string in $AdS_5\times S^5$ the WS method gives frequencies with integer mode numbers \cite{wssl2}, while the AC gives non-trivial shifts \cite{GV07}. The mode numbers are also taken integer in \cite{SchaferNameki:2005tn, HL}. For the $su(2)$ string in $AdS_5\times S^5$ the WS can give half-integer mode numbers for one form of solution \cite{wssu2} and integer mode numbers for another \cite{rform}\footnote{We thank A.~Tseytlin for pointing out this disagreement to us.}. The mode numbers are also taken integer in \cite{Beisert:2005mq, HL} and half-integer in \cite{Frolov:2004bh, SchaferNameki:2006gk}. In $AdS_4\times\CP^3$ for the $su(2)$ string the WS and the AC results agree \cite{Chuvaki}, while for the $sl(2)$ the half-integer shifts for fermions are again different \cite{McLRTs, NG}. 

The half-integer shifts in the mode numbers for fermionic modes are related to the fermionic boundary conditions (f.b.c.): switching from periodic to antiperiodic fermions produces an extra $1/2$ shift in the mode numbers. The discrepancies listed above suggest that the f.b.c. in the WS approach are being chosen somehow improperly. If the target space were not simply connected, the f.b.c. for the string that winds around a non-trivial cycle indeed could be chosen either periodic or antiperiodic, depending on the spin structure. It corresponds to the fact that in the spin bundle the two edges of the chart that wraps the cycle can be glued either with matrix ${\bf 1}$ or ${\bf -1}$ from the $Spin$ group. Changing periodic to antiperiodic f.b.c. for a string that winds $m$ times around the cycle would produce a shift $m/2$ in the mode numbers~---~exactly what we need!

Thus the idea is to construct carefully the spin bundle over the target space. In the WS calculations the target space is usually parametrized by angular coordinates. They are defined in a ``big'' chart, but the neighborhoods of the coordinate singularities must be glued up with small extra patches with Cartesian coordinates. The ``big'' chart is not simply connected, and thus has several spin structures, but only one of them comes from the restriction of the spin bundle over the whole target space.
We prove that the spin structure that can be continued from the ``big'' chart to the whole space is the one that has transition matrices ${\bf -1}$ on the cycles of the ``big'' patch. The f.b.c. corresponding to the correct spin structure give frequencies that do agree with the algebraic curve.

One way to avoid these complications with spin structures is to start with the coset form of the GS action, where one does not need to choose any particular coordinates on the target space. For $AdS_5\times S^5$ it has been checked indeed (see App. E of \cite{GV07}) that the frequencies for circular strings, computed starting from the coset action, agree with the algebraic curve without any further modifications. For the string in $AdS_4\times\CP^3$ it has been checked \cite{GMV} that the GS action of \cite{McLRTs} can be obtained from the coset action after a {\it non-periodic} rotation of fermions, which indicates problems with periodicity.

For the $sl(2)$ circular string in $AdS_4\times\CP^3$ there is also a disagreement between WS \cite{McLRTs} and AC \cite{NG} for the {\it bosonic} $\CP$ light modes. In the paper we explain that this disagreement is due to the specific choice of the classical solution in \cite{McLRTs}, which becomes non-periodic when a fluctuation is added to some of the string coordinates.

The paper is organized as follows. In section 2 we explicitly construct the spin bundles for manifolds of interest. Sections 3 and 4 contain the discussion of fermionic frequencies for the circular strings in $AdS_5\times S^5$ and $AdS_4\times \CP^3$, respectively. In section 5 we present our conclusions. The discussion of the $\CP$ bosonic frequency for the $sl(2)$ string is presented in the Appendix.

\section{Spin bundles and angular coordinates}
\begin{floatingfigure}[p]{70mm}
        \centering
        \includegraphics[width=70mm]{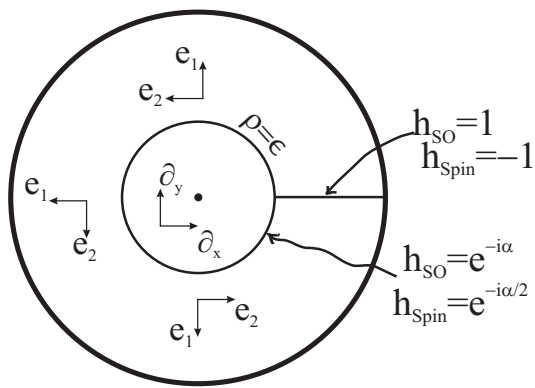}
        \label{fig}
\vspace{-4mm}
\caption{\small Charts and transition functions for the tangent and spin bundles over disk with polar coordinates.}
\vspace{2mm}
\end{floatingfigure}
\sloppy Recall \cite{DeWitt} that a vector bundle over a manifold can be defined by specifying a set of charts covering the manifold, plus a system of transition functions on the intersections. For the tangent bundle the orthonormal frame in each chart is the vielbein, and the transitions matrices $h_{SO}(t)$ are valued in $SO(n)$. For the spin bundle the transition matrices $h_{Spin}(t)$ are valued in $Spin(n)$, which is a double cover of the $SO(n)$ group, $Spin(n)\xrightarrow{~p~}SO(n)$. If the spin bundle comes from the reduction of the tangent bundle, then its transition matrices cover in each point the transition matrices of the tangent bundle, $p\left\{h_{Spin}(t)\right\}=h_{SO}(t)$.

\fussy For a spin manifold $M$ the spin structures are classified by homomorphisms $Hom(\pi_1(M),{\mathbb Z}_2)$, {\it i.e.} for each fundamental cycle we can choose transition matrix ${\bf 1}$ or ${\bf -1}\in Spin(n)$ (equivalently, R or NS boundary conditions for the fermions). For a simply connected spin manifold the spin structure is unique.

As an example consider a unit two dimensional disk (see fig.~1). In standard polar coordinates $(\rho, \A)$ the metric and the vielbein are
\beqn
&&ds^2=d\rho^2+\rho^2d\A^2\,,\nnr
&&e_1=\pt_\rho\,,\,\,e_2=\fr{1}{\rho}\pt_\A\,.
\eeqn
The polar coordinates are good only in the cylinder $\eps<\rho<1$, which we call the $(\rho, \A)$-chart. On the two ends of this chart, $\A=0$ and $\A=2\pi$, the frames in the tangent bundle are glued via identity matrix. The vicinity of the origin, $\rho<\eps$, must be covered with another chart\footnote{For simplicity we write for the charts $\rho<\eps$ or $\rho>\eps$, but of course it is implied that the charts are a bit larger, so that they overlap.} with flat coordinates $x$, $y$. On the intersection the vielbeins $(e_1, e_2)$ and $(\pt_x,\pt_y)$ are glued via an $SO(2)\approx U(1)$ transformation, which in complex notations acts as
\beq
e_1+ie_2=h_{SO}\cdot(\pt_x+i\pt_y)\,,\,\,h_{SO}(\rho,\A)=\ex^{-i\A}\,.
\eeq
In two dimensions $Spin(2)\approx U(1)$, the projection $p$ acts as $p\left\{h_{Spin}\right\}=\left(h_{Spin}\right)^2$. For our transition function $h_{SO}(\rho,\A)$ the spin covering is 
\beq
h_{Spin}(\rho,\A)=\sqrt{h_{SO}(\rho,\A)}=\ex^{-i\A/2}\,,
\eeq
which is not periodic. It means that the $\A=0$ and $\A=2\pi$ ends of the $(\rho, \A)$-chart must be glued with transition function $({\bf -1})\in Spin(2)$ for the spin bundle.

We would like to stress again that on the cylinder we are free to take any of the two spin structures ({\it i.e.}, periodic or antiperiodic fermions). Examination of how the endcap is glued to the cylinder allows us to choose the unique spin structure, which is realized on the simply connected disk.

Now let us consider how the spin structure is realized on the odd-dimensional sphere in angular coordinates. For instance, take $S^5$. The coordinates are chosen as
\beqn
&&z_1=\rho_1\ex^{i\alpha_1}\,,\nnr
&&z_2=\rho_2\ex^{i\alpha_2}\,,\\
&&z_3=\rho_3\ex^{i\alpha_3}\,,\nonumber
\eeqn
where $\sum \rho_i^2=1$, $\rho_i\ge0$ and in principle $\rho_i$ are expressed in some way in terms of two angular coordinates $\theta_1, \theta_2$. This coordinate system is good in the ``big'' chart $\rho_{1,2,3}>\eps$ which has topology $T^3\times I_2$, where $I_2$ is a triangle. The rest of $S^5$ is covered by six intersecting patches, where the pairs $(\rho_i,\A_i)$, for which $\rho_i<\eps$, are replaced by Cartesian coordinates $x_i=\rho_i\cos\A_i, y_i=\rho_i\sin\A_i$.

The ``big'' chart is not simply connected, and there can be $2^3=8$ spin structures on it, but only one of them is realized as a restriction of the unique $S^5$ spin structure. To fix it we consider the intersection\footnote{More exactly, the slice of this intersection for constant coordinates $\rho_2, \A_2, \A_3$.} of the ``big'' chart with the chart $\rho_1<\eps, \rho_{2,3}>\eps$ parameterized by $\rho_2, \A_2, \A_3$ and Cartesian coordinates $x_1, y_1$. The coordinates $\rho_2, \A_2, \A_3$ are the same on the two sides of the intersection, and can be safely forgotten~---~the transition matrix acts trivially in these directions. Thus we are exactly in the situation of our previous example with the disk. The frame $(\pt_{\rho_1}, \fr{1}{\rho_1}\pt_{\alpha_1})$ rotates on $2\pi$ w.r.t. the frame $(\pt_{x_1}, \pt_{y_1})$ when going around the intersection of two patches. Since this rotation is a non-trivial cycle in $SO(5)$, it cannot be covered by a cycle in $Spin(5)$ and an additional cut in the ``big'' chart is required, where the transition matrix for the spin bundle is ${\bf -1}\in Spin(5)$. This cut goes along the hypersurface $\alpha_1={\text const}$. A little subtlety is that this argument deals with the frame $(\pt_{\rho_1}, \fr{1}{\rho_1}\pt_{\alpha_1}, \pt_{\rho_2}, \pt_{\A_2}, \pt_{\A_3})$, which is not the vielbein. But for $\rho_i={\text const}$ this frame is related to the vielbein simply by multiplication by a constant matrix.

In the same way there are cuts along the hypersurfaces $\A_2={\text const}$ and $\A_3={\text const}$, where the trivial transition matrix ${\bf 1}\in SO(5)$ is covered by matrix ${\bf -1}\in Spin(5)$. Thus we have found the spin structure on the ``big'' patch which comes from the unique spin structure on $S^5$.

Another example which is important for us is $\CP^n$. Let the coordinates be
\beqn
&z_1=\rho_1\ex^{i\alpha_1}\,,&\nnr
&\dots&\nnr
&z_n=\rho_n\ex^{i\alpha_n}\,,&\label{CPcoord}\\
&z_{n+1}=\rho_{n+1}\,,& \nonumber
\eeqn
where $\sum \rho_i^2=1$ and we have also fixed the phase of $z_{n+1}$ to be zero. Just as in the case of sphere one can see that for the spin bundle there are cuts in the ``big'' chart along the hypersurfaces $\alpha_i={\text const}$ with transition matrix ${\bf -1}\in Spin(2n)$. Considering the intersection of the ``big'' chart with the chart $\rho_{n+1}<\eps$ we find that there should also be a cut in the cycle parameterized by $\sum_{i=1}^n\alpha_i$, or by $\alpha_{n+1}$ in another gauge. It implies the equation $(-1)^n=-1$, {\it i.e.} that $n$ is odd. But that is just the well known fact that the projective space $\CP^n$ is not spin for $n$ even.\\
~\\
\section{Circular strings in $AdS_5\times S^5$}
We take the angular coordinates as in \cite{wssu2},
\beqn
&&Y_0=\cosh\rho\,\ex^{it}\,,\quad Y_1=\sinh\rho\sin\theta\ex^{i\vphi_1}\,,\quad Y_2=\sinh\rho\cos\theta\ex^{i\vphi_2}\,,\nnr
&&X_1=\sin\gamma\cos\psi\ex^{i\phi_1}\,,\quad X_2=\sin\gamma\sin\psi\ex^{i\phi_2}\,,\quad X_3=\cos\gamma\ex^{i\phi_3}\,,
\eeqn
where $Y_i$ and $X_i$ are the embedding coordinates for $AdS_5$ and $S^5$ respectively. The angular coordinates vary in the range $0\le\psi\le \pi/2$, $0\le\theta\le \pi/2$, $0\le\gamma\le \pi/2$, $0\le\phi_{1,2}<2\pi$, $0\le\vphi_{1,2}<2\pi$.

\subsection {The $su(2)$ circular string}
For the $su(2)$ circular string two forms of the solution have been considered in the literature \cite{wssu2, rform}. For both of them $t=\kappa\tau$, $\rho=0$, $\gamma=\pi/2$, $\phi_3=\nu\tau$, and the other coordinates are
\beqn
\bea{llll}
\text {first form \cite{rform}:}&\psi=\pi/4\,,  &\phi_1=w\tau+m\sigma\,,&\phi_2=w\tau-m\sigma\,,\\
\text {second form \cite{wssu2}:}&\psi=m\sigma\,,&\phi_1=w\tau\,,        &\phi_2=w\tau\,.
\eea
\eeqn
The two forms are related by an $SO(6)$ rotation in $S^5$, but amazingly the fermionic frequencies for them obtained by explicit semiclassical computation are different. For the second form the mode numbers of the frequencies are half-integer \cite{wssu2}, while for the first form they are integer (see tab.\ref{tab1}\footnote{Note that all the formulae in our paper are valid only up to integer shifts in the mode-numbers, which we do not consider here. We also ignore the constant shifts of the frequencies.}), as one can easily check along the lines of \cite{wssu2}. It is natural to expect that this discrepancy is some coordinate artefact. Indeed we show that it disappears after taking into account the transition functions of the spin bundle. 

\begin{table}[h]
\caption{\la{tab1} \small Fermionic frequencies for the circular strings in $AdS_5\times S^5$ according to worldsheet $\rm{(ws)}$ \cite{wssu2,rform,wssl2} and algebraic curve $\rm{(ac)}$ \cite{GV07} calculations. Here $\rm{ws_{1}}$ and $\rm{ws_2}$ denote $\rm{ws}$-calculations for the first and the second form of the $su(2)$ solution, respectively. $\rm{\widetilde{ws}}$ denotes the worldsheet result with our modification of fermionic periodicity taken into account. For the notations see tab.\ref{tabnot}. Our expressions are correct up to constant shifts of the frequencies and integer shifts in the mode numbers.}
\beq \nn
\bea{c|l}
\toprule
su(2)
&
\bea{c|c|c|c|c}
{\rm\bf ~ws_1~}&{\rm\bf ws_2}&{\rm\bf~~~~~~~\,\widetilde{ws}~~~~~~~}&{\rm\bf ~~~~~~~~ac~~~~~~~}&{\rm\bf multiplicity}\\
\midrule
\omega^{F_1}_{n} & \omega^{F_1}_{n+m/2}& \omega^{F_1}_n& \omega^{F_1}_{n} & \times 8
\eea
\\
\midrule
sl(2)
&
\bea{c|c|c|c}
{\rm\bf ~~~~~~~ws~~~~~~~~}&{\rm\bf~~~~~~\widetilde{ws}~~~~~}&{~~~~~\rm\bf ac~~~~~}&{\rm\bf multiplicity}\\
\midrule
\bea{c} \omega^{F_2}_{n}\\ \omega^{F_2}_{-n} \eea & \bea{c} \omega^{F_2}_{n+m/2+k/2}\\ \omega^{F_2}_{-n+m/2+k/2}\eea & \bea{c} \omega^{F_2}_{n+m/2-k/2}\\ \omega^{F_2}_{-n-m/2-k/2}\eea & \bea{c} \times 4\\\times 4\eea
\eea
\\
\bottomrule
\eea
\eeq
\end{table}

In what follows we take $\tau=0$ for simplicity. In both forms the string lies in the coordinate singularities for the $AdS$ angular coordinates and $\phi_3$ in $S^5$. To make sense of the solution we deform it slightly, $\gamma\rightarrow \pi/2+\eps_1$, $\rho\rightarrow 0+\eps_2$, moving it to the region with well defined vielbein.

According to the result of the previous section, in the ``big'' chart on $S^5$ there are cuts along three hypersurfaces $\phi_{1,2,3}={\text const}$, on which the spin bundle has transition matrix ${\bf -1}$. For a string crossing such a cut once the boundary conditions for the fermions should be changed to antiperiodic. The solution in the first form crosses the cut in $\phi_1$ $m$ times and the cut in $\phi_2$ $m$ times. Hence for any integer $m$ the fermionic boundary conditions remain periodic, and the naive answer for the fluctuation frequencies is correct.

For the second form of solution the problem is more complicated. The string does not cross the cuts, but it passes just through the $\phi_1$ and $\phi_2$ coordinate singularities, and it cannot be moved from the singularities by a small deformation of the solution. Also, for this solution it is assumed that $\psi$ can run from 0 to $2\pi$, and it means that the associated vielbein is different from what we have chosen. In principle one can construct the system of patches such that $\psi$ can run up to $2\pi$, but we choose to take the solution in the form where $\psi$ does not run out of the $[0,\pi/2]$ interval. Then the string is divided into four parts (we take $m=1$),
\beqn
\begin{array}{c|clcc}
 &\sigma                 &\psi        &\phi_1 &\phi_2\\\hline
1&0\rightarrow\pi/2      &\sigma      &0      &0\\
2&\pi/2\rightarrow\pi    &\pi-\sigma  &\pi    &0\\
3&\pi\rightarrow 3\pi/2  &\sigma-\pi  &\pi    &\pi\\
4&3\pi/2\rightarrow 2\pi &2\pi-\sigma &2\pi   &\pi
\end{array}\nonumber
\eeqn

In what follows we revisit the calculation of fermionic frequencies of \cite{wssu2}, closely following the notations of that paper. After the $\kappa$-symmetry fixing the fermionic part of the Lagrangian is
\beqn
&&L_F=2i\bar\theta D\theta\,,\label{LagrAdS}\\
&&D=-\rho^a D_a+\fr{1}{2}\eps^{ab}\rho_a\Gamma_{01234}\rho_b\,,\nnr
&&D_a=\pt_a+\fr14\omega_a^{AB}\Gamma_{AB}\,,\quad \rho_a=\pt_aX^\mu e_\mu^A \Gamma_A\,,\quad \omega_a^{AB}=\pt_aX^\mu\omega_\mu^{AB}\,,
\eeqn
where $e_\mu^A$ is the vielbein and $\omega_\mu^{AB}$ is the spin connection.

The boundary conditions for $\theta$ on the transitions between regions 1,2,3,4 are non-trivial, because the vielbein is rotated by $\pi$ when a singularity is crossed. More careful inspection shows that the following boundary conditions should be imposed,
\beqn
\theta_2(\pi/2) &=&\ex^{\fr{\pi}{2}\Gamma_{78}}~\theta_1(\pi/2)\,,\nnr
\theta_3(\pi)   &=&\ex^{\fr{\pi}{2}\Gamma_{79}}~\theta_2(\pi)\,,\nnr
\theta_4(3\pi/2)&=&\ex^{-\fr{\pi}{2}\Gamma_{78}}~\theta_3(3\pi/2)\,,\nnr
\theta_1(2\pi)  &=&\ex^{-\fr{\pi}{2}\Gamma_{79}}~\theta_4(2\pi)\,,\nonumber
\eeqn
where the subscripts 1,2,3,4 label the region where $\theta$ is taken. For example, these boundary conditions say that upon crossing the singularity between regions 1 and 2 the frame is rotated by~$-\pi$ in the $78$-plane, and hence the variable $\theta$ should be rotated by $\pi$. The direction of rotation is important since a $2\pi$ rotation of spinors produces minus one. The directions can be unambiguously defined only after the cuts in the ``big'' chart have been chosen.

Upon substituting everything to the eq. (\ref{LagrAdS}), the operator $D$ takes the form
\beq
D=(\kappa\Gamma_0+w\Gamma_{\tilde 8})(\pt_\tau-\fr12\nu\Gamma_{65}+\fr12w\Gamma_{\tilde{9}7})-s\Gamma_7\pt_\sigma+sw\Gamma_7\Gamma_{\tilde{8}}\Gamma_{01234}\,,
\eeq
where $s=\pt_\sigma\psi=\pm1$ and
\beq
\Gamma_{\tilde{8}}=\Gamma_8\cos\psi+\Gamma_9\sin\psi\,,\quad \Gamma_{\tilde{9}}=\Gamma_9\cos\psi-\Gamma_8\sin\psi\,.
\eeq
To remove the $\sigma$ dependence that enters the equation through $\psi(\sigma)$ we apply a rotation in the $89$-plane,
\beq
\theta=\ex^{-\fr12\psi\Gamma_{89}}\tilde{\theta}\,,
\eeq
After that the operator takes the form
\beq
\tilde{D}=(\kappa\Gamma_0+w\Gamma_8)(\pt_\tau-\fr12\nu\Gamma_{65}+\fr12w\Gamma_{97})-s\Gamma_7\pt_\sigma+\fr12\Gamma_7\Gamma_{89}+sw\Gamma_7\Gamma_{\tilde{8}}\Gamma_{01234}\,.
\eeq
The boundary conditions change as follows,
\beqn
\tilde{\theta_2}(\pi/2) &=&\ex^{\fr{\pi}{4}\Gamma_{89}}~\ex^{\fr{\pi}{2}\Gamma_{78}}~\ex^{-\fr{\pi}{4}\Gamma_{89}}~\tilde{\theta_1}(\pi/2)\,,\nnr
\tilde{\theta_3}(\pi)   &=&\ex^{\fr{\pi}{2}\Gamma_{79}}~\tilde{\theta_2}(\pi)\,,\nnr
\tilde{\theta_4}(3\pi/2)&=&\ex^{\fr{\pi}{4}\Gamma_{89}}~\ex^{-\fr{\pi}{2}\Gamma_{78}}~\ex^{-\fr{\pi}{4}\Gamma_{89}}~\tilde{\theta_3}(3\pi/2)\,,\nnr
\tilde{\theta_1}(2\pi)  &=&\ex^{-\fr{\pi}{2}\Gamma_{79}}~\tilde{\theta_4}(2\pi)\,.\nonumber
\eeqn
After one more change of variables
\beqn
\tilde{\theta_1} &=&\hat{\theta_1}\,,\nnr
\tilde{\theta_2} &=&\ex^{-\fr{\pi}{2}\Gamma_{79}} \hat{\theta_2}\,,\nnr
\tilde{\theta_3} &=&\hat{\theta_3}\,,\nnr
\tilde{\theta_4} &=&\ex^{\fr{\pi}{2}\Gamma_{79}}\hat{\theta_4}\nonumber
\eeqn
the alternating sign $s$ disappears from $\tilde{D}$, and the boundary conditions become trivial periodic, giving us frequencies with integer mode numbers. We conclude that after taking into account the transition matrices the agreement between two worldsheet calculations of frequencies is restored, and they also agree with the algebraic curve result (see tab.\ref{tab1}).

\subsection {The $sl(2)$ circular string}
The next example is the $sl(2)$ circular string. The classical string motion is described by \cite{wssl2}
\beqn
&&t=\kappa\tau\,,\quad \rho=\rho_0\,,\quad\theta=\pi/2\,,\quad \vphi_1=w\tau+k\sigma\,,\nnr
&&\gamma=\pi/2\,,\quad \psi=0\,,\quad \phi_1=w\tau-m\sigma\,.
\eeqn
For the spin bundle over $AdS_5$ similarly to the sphere there are cuts along the hyperplanes $\varphi_1={\text const}$ and $\varphi_2={\text const}$. After a small deformation of the solution, $\gamma\rightarrow \pi/2+\eps$, the string crosses the $\varphi_1$ cut $k$ times and the $\phi_1$ cut $m$ times. Hence for even $k+m$ the naive result for the frequencies is correct, while for odd $k+m$ the mode numbers should be shifted by $1/2$. Up to integer shift the result can be concisely written as $\omega_{n+k/2+m/2}$, where $\omega_n$ is the naive result without cuts. Note that the corrected answer agrees with the algebraic curve (tab.\ref{tab1}).

\section{Circular strings in $AdS_4\times \CP^3$}
\subsection {The $su(2)$ circular string}
The $su(2)$ circular string in $AdS_4\times\CP^3$ has been considered in a recent paper \cite{Chuvaki}. In that paper the following parametrization of the homogenious coordinates is taken,
\beqn
&&Z_1=\cos\xi\cos\fr{\theta_1}{2}\ex^{i(\psi+\phi_1)/2}\,,\nnr
&&Z_2=\cos\xi\sin\fr{\theta_1}{2}\ex^{i(\psi-\phi_1)/2}\,,\nnr
&&Z_3=\sin\xi\cos\fr{\theta_2}{2}\ex^{i(-\psi+\phi_2)/2}\,,\nnr
&&Z_4=\sin\xi\sin\fr{\theta_2}{2}\ex^{i(-\psi-\phi_2)/2}\,.\nonumber
\eeqn
The metric is
\beqn
ds^2_{AdS_4}&=&-\cosh^2\rho~dt^2+d\rho^2+\sinh^2\rho~(d\theta^2+\sin^2\theta d\vphi^2)\,,\label{metr1}\\
ds^2_{\CP^3}&=&d\xi^2+\cos^2\xi\sin^2\xi\left(d\psi+\fr12\cos\theta_1d\phi_1-\fr12\cos\theta_2d\phi_2\right)^2\nnr
&&+\fr14\cos^2\xi(d\theta_1^2+\sin^2\theta_1d\phi_1^2)+\fr14\sin^2\xi(d\theta_2^2+\sin^2\theta_2d\phi_2^2)\,.\nonumber
\eeqn
The circular string solution looks like
\beqn
&&t=\kappa\tau\,,\quad \rho=0\,,\nnr
&&\xi=\pi/4\,,\quad \theta_1=\theta_2=\pi/2\,,\quad \psi=m\sigma\,, \phi_1=\phi_2=\omega\tau\,.
\eeqn
It is tempting to say that there is a cut along $\psi={\text const}$, which the string crosses $m$ times, thus producing an extra shift $m/2$ in the mode numbers. But that would be a wrong answer. First it is necessary to fix the range of variation of the angular coordinates $\psi$, $\phi_1$, $\phi_2$. The minimal periods for them are
\beqn
&&\Delta\psi=(n_1+n_2-n_3)\pi\,,\nnr
&&\Delta\phi_1=2(n_1-n_2)\pi\,,\nnr
&&\Delta\phi_2=2n_3\pi\,,\quad n_{1,2,3}\in{\mathbb Z}\,.
\eeqn
Of course, these coordinates are separately periodic with periods $2\pi$ for $\psi$ and $4\pi$ for $\phi_{1,2}$, but these are not the minimal periods. These angles are related to the manifestly periodic angles of (\ref{CPcoord}) as follows,
\beqn
&&\psi=(\A_1+\A_2-\A_3)/2\,,\nnr
&&\phi_1=\A_1-\A_2\,,\nnr
&&\phi_2=\A_3\,,\
\eeqn
Hence on the solution $\A_1=m\sigma$ and $\A_2=m\sigma$ (as usual, here $\tau=0$). The string crosses the $\A_1$ cut and the $\A_2$ cut $2m$ times in total, thus there are no additional half-integer shifts in the fermionic mode numbers, and the worldsheet result of \cite{Chuvaki} is correct.

\subsection {The $sl(2)$ circular string}
The $sl(2)$ circular string in $AdS_4\times\CP^3$ has been considered in \cite{McLRTs}. The coordinates used there are:
\beqn
&&Z_1=\ex^{i\A_1}\cos\zeta_1\,,\nnr
&&Z_2=\ex^{i\A_2}\sin\zeta_1\cos\zeta_2\,,\nnr
&&Z_3=\ex^{i\A_3}\sin\zeta_1\sin\zeta_2\cos\zeta_3\,,\nnr
&&Z_4=\sin\zeta_1\sin\zeta_2\sin\zeta_3\,,\nonumber
\eeqn
$Z_i$ being the homogenious coordinates. The variables $\A_i$ are $2\pi$-periodic, and $\zeta_i$ vary from $0$ to $\pi/2$\footnote{Our angles are related to the variables of \cite{McLRTs} as follows, $\A_1=\tau_1+\tau_2+\tau_3=\phi_2+\phi_3$, $\A_2=\tau_2+\tau_3=\phi_3+\phi_1$, $\A_3=\tau_3=\phi_1+\phi_2$.}. 
The $AdS$ coordinates are as in the previous subsection. The $\CP$ metric is 
\beqn
ds^2_{\CP}&=&d\zeta_1^2+\sin^2\zeta_1\left[d\zeta_2^2+\cos^2\zeta_1(d\A_1-\cos^2\zeta_2d\A_2-\sin^2\zeta_2\cos^2\zeta_3d\A_3)^2\right.\nnr
     &&\left.+\sin^2\zeta_2(d\zeta_3^2+\cos^2\zeta_2(d\A_2-\cos^2\zeta_3d\A_3)^2+\sin^2\zeta_3\cos^2\zeta_3d\A_3^2)\right]\,.\label{metr2}
\eeqn
For the solution considered in \cite{McLRTs}
\beqn
&&t=\kappa\tau\,,\quad \rho=\text{const}\,,\quad\theta=\fr{\pi}{2}\,,\quad \vphi=w\tau+k\sigma\,,\nnr
&&\zeta_1=\fr{\pi}{4}\,,\quad \zeta_2=\zeta_3=\fr{\pi}{2}\,,\quad \A_1=\omega\tau+m\sigma\,,\quad \A_2=\A_3=\fr12(\omega\tau+m\sigma)\,. \label{sl1}
\eeqn
It appears that this solution is problematic. The submanifold $\zeta_2=\zeta_3=\fr{\pi}{2}$ is singular for $\A_2$ and $\A_3$, and it is impossible to move the string from this singularity, because it becomes non-periodic after the deformation $\zeta_{2,3}\rightarrow\pi/2+\eps$. To fix the problem we take another solution that differs from this one only by the values of singular coordinates, {\it i.e.} describes the same motion of the string in the target space, but can be moved from the singularity. We set
\beq
\zeta_1=\fr{\pi}{4}\,,\quad \zeta_2=\zeta_3=\fr{\pi}{2}\,,\quad \A_1=\omega\tau+m\sigma\,,\quad \A_2=\A_3=0\,.\label{sl2}
\eeq
Actually, this is an example of a general phenomenon. For any circular string the $\tau$ and $\sigma$ dependence of the singular angular coordinates can be taken arbitrary, because these coordinates drop out of the equations of motion and Virasoro constraints. But this dependence does enter the fermionic Lagrangian through the spin connection.\footnote{One can see that the dependence of a singular coordinate on $\tau$ changes the constant shifts in the frequencies, and the dependence on $\sigma$ changes the shifts in the mode numbers. The $AdS_5\times S^5$ $su(2)$-string and the $AdS_4\times\CP^3$ $sl(2)$-string provide the examples.} To fix the ambiguities one should take a more general solution that generically does not lie on the singularity and then take the limit. For example, recall our discussion of the $su(2)$ string in $AdS_5\times S^5$. The coordinate $\phi_3=\nu\tau$ for that string is singular when $\gamma=\pi/2$, and hence the parameter $\nu$ is not fixed by the equations of motion, but it does enter the Lagrangian like $\pt_0-\fr12\nu\Gamma_{65}+...$.  The correct value $\nu=\sqrt{w^2-m^2}$ is fixed after one takes a more general solution with $\gamma\ne\pi/2$, see \cite{wssu2}. In the same way for the $sl(2)$ string in $AdS_4\times\CP^3$ we can justify our choice of $\A_{2,3}$: for $\A_2=\A_3=0$ there exists solution with arbitrary values of $\zeta_2$ and $\zeta_3$, which is just a symmetry transformation of the solution (\ref{sl2}).

We have repeated the computation of \cite{McLRTs} for the modified solution (\ref{sl2}). Essentially all the difference from the computation of that paper is that in our case the $\CP$ part of the spin connection is non-trivial. The projected spin connection is equal to
\beqn
&&\omega_0=2\kappa\sinh\rho\G_{01}+2w\cosh\rho\G_{31}+\omega(\G_{96}-\G_{58})\,,\nnr
&&\omega_1=2k\cosh\rho\G_{31}+m(\G_{96}-\G_{58})\,,
\eeqn
where the coordinates are enumerated as $(t,\rho,\theta,\vphi)\equiv(x_0,x_1,x_2,x_3)$ and $(\zeta_1,\zeta_2,\zeta_3,\A_1,\A_2,\A_3)\equiv(x_4,x_5,x_6,x_7,x_8,x_9)$. Repeating the same steps as in \cite{McLRTs}, we arrive at the result for the frequencies that is stated in tab.\ref{tab2}. As expected, the different choice for the behaviour of the singular coordinates has resulted in the shifts of the mode numbers, compared to \cite{McLRTs}.

Now the transition matrices of the spin bundle must be taken into account. The string crosses the $\A_1$ cut in $\CP^3$ $m$ times, hence $m/2$ has to be added to the mode numbers. In the same way a shift by $k/2$ comes from the cut in $\vphi$ in the $AdS_4$. Again, the resulting frequencies agree with the algebraic curve (see tab.\ref{tab2}).

\begin{table}[h]
\caption{\la{tab2} \small Fermionic frequencies for circular strings in $AdS_4\times\CP^3$. The frequencies in the columns $\rm{ws}_1$ and $\rm{ws}_2$ correspond to the solution (\ref{sl1})\cite{McLRTs} and (\ref{sl2}), respectively. The column ${\rm ac}$ presents the result of the algebraic curve computation \cite{NG}. In the column $\rm{\widetilde{ws}}$ our modification of fermionic periodicity is taken into account. For the notations see tab.\ref{tabnot}. Our expressions are correct up to constant shifts of the frequencies and integer shifts in the mode numbers.}
\beq \nn
\bea{c|l}
\toprule
su(2) & 
\bea{c|c|c|c}
{\rm\bf ~~~~~~~~~~~~~~ws~~~~~~~~~~~~~} & {\rm\bf~~~~~~~~\widetilde{ws}~~~~\,~~} & {\rm\bf~~~~~~~~ac~~~~~~~} & {\rm\bf multiplicity}\\
\midrule
\bea{c} \omega^{F_1}_{2n}/2\\\omega^{A}_n \eea & \bea{c} \omega^{F_1}_{2n}/2\\\omega^{A}_n \eea & \bea{c} \omega^{F_1}_{2n}/2\\ \omega^{A}_n \eea & \bea{c} \times4\\\times4\eea
\eea \\
\midrule
sl(2) &
\bea{c|c|c|c|c}
{\rm\bf ws_1} & {\rm\bf ws_2} & {\rm\bf\widetilde{ws}} & {\rm\bf ac} & {\rm\bf multiplicity}\\
\midrule
\bea{c}
 \omega^{F_2}_n\\
 \omega^{F_2}_{-n}  \\
 \omega^{A_+}_{2n}/2  \\
 \omega^{A_-}_{2n}/2  \\      
\eea
&
\bea{c}
 \omega^{F_2}_n\\
 \omega^{F_2}_{-n}  \\
 \omega^{A_+}_{2(n+m/2)}/2  \\
 \omega^{A_-}_{2(n+m/2)}/2  \\      
\eea
& 
\bea{c}
\omega^{F_2}_{n+m/2+k/2} \\
\omega^{F_2}_{-n+m/2+k/2}  \\
\omega^{A_+}_{2(n+k/2)}/2 \\
\omega^{A_-}_{2(n+k/2)}/2 \\
\eea
&
\bea{c}
\omega^{F_2}_{n+m/2-k/2}\\ 
\omega^{F_2}_{-n-m/2-k/2}\\
\omega^{A_+}_{2(n+k/2)}/2    \\
\omega^{A_-}_{2(n-k/2)}/2    
\eea
&
\bea{c}
\times 2\\
\times 2\\
\times 2\\
\times 2
\eea
\eea
\\
\bottomrule
\eea
\eeq
\end{table}

\begin{table}[h]
\caption{\la{tabnot} \small Notations for the frequencies.}
\beq \nn
\bea{c|c}
\toprule
{\rm\bf notation} & {\rm\bf frequency}\\
\midrule
\omega^{F_1}_n & \sqrt{n^2+\omega^2}\\
\midrule
\omega^{F_2}_n & \sqrt{\left(n+\fr{\sqrt{w^2-\omega^2}}{2}\right)^2+\fr12(\kappa^2+\omega^2-m^2)}\\
\midrule
\omega^{A}_n & \sqrt{n^2+\kappa^2}\\
\midrule
\omega^{A_\pm}_n & x:~~(x^2-n^2)^2+4\sinh^2\rho\kappa^2x^2-4\cosh^2\rho(wx-kn)^2=0\\
\midrule
\omega^S_n & \sqrt{n^2+\omega^2-m^2}\\
\bottomrule
\eea
\eeq
\end{table}

\section{Conclusions}
In this paper we have revisited the computation of semiclassical frequencies of circular strings in $AdS_5\times S^5$ and $AdS_4\times \CP^3$, starting from the Green-Schwarz action. We have found that in order to fix the correct fermionic boundary conditions one has to construct accurately the system of transition matrices for the spin bundle over the target space. After taking into account the transition matrices the modified worldsheet results agree with the known algebraic curve frequencies (up to integer part of the mode number shifts, which we do not discuss here). We have explicitly considered simple $su(2)$ and $sl(2)$ strings, but our findings apply to any classical string that cannot be contracted without crossing the coordinate singularities.

We have also explained a minor disagreement between the AC and WS computations of {\it bosonic} spectrum for the $sl(2)$ string living in $AdS_4\times\CP^3$. We point out that one has to be especially careful dealing with strings that lie in singularity of some of the angular coordinates. Also, all the problems discussed here do not appear in computations based on the coset form of the string action.

\section*{Acknowledgments} I am grateful to N.~Gromov, T.~McLoughlin, S.~Oblezin, R.~Roiban, K.~Zarembo, P.~Vieira and especially A.~Tseytlin for many useful discussions. I also thank Imperial College London for hospitality. I am supported by the RFBR grant 09-02-00308-a, the grant PICS-07-0292166, the grant for support of scientific schools NSh-3036.2008.2, and by the Dynasty foundation.

\section*{Appendix. Bosonic $\CP$ frequency}
In this Appendix we comment further on the discrepancies between the worldsheet calculation of \cite{McLRTs} and the algebraic curve result \cite{NG}. The difference in fermionic frequencies has been fixed in sec.~4, up to matters of regularization, but it appears that half-integer shifts in some bosonic frequencies also do not agree, namely in the four ``light'' $\CP$ modes (see tab.\ref{tabbos}). The origin of this disagreement is the following. As we have explained in sec.~4, the solution (\ref{sl1}) is problematic for calculation of the fermionic frequencies. For the bosonic frequencies the problem with (\ref{sl1}) is even more obvious: though it is periodic itself, it becomes non-periodic when a fluctuation in $\zeta_2$ or $\zeta_3$ is excited. We have calculated the bosonic frequencies for the corrected solution (\ref{sl2}), and they do agree with the algebraic curve (see tab.\ref{tabbos}). This computation is quite standard, so we do not present it here. One starts with the bosonic Lagrangian
\beq
{\cal L}_{bos}=g^{AdS}_{\mu\nu}\pt_a y^{\mu}\pt_a y^{nu}+4g^{\CP}_{\mu\nu} \pt_a x^{\mu}\pt_a x^{\nu}\,,\label{actbos}
\eeq
where the factor $4$ comes from the fact that the $\CP$ radius is twice the radius of $AdS$.
The metric is as in equations (\ref{metr1}) and (\ref{metr2}). The coordinates $x^\mu$ and $y^\mu$ are the $\CP$ and $AdS$ angular coordinates. We expand the action (\ref{actbos}) to the quadratic order near the solution (\ref{sl2}). The frequencies are then read off from the determinant of the matrix that appears in the linearized equations of motion. 

However, there is a little subtlety. The coordinates $\A_2$ and $\A_3$ are singular on the solution (\ref{sl2}), hence their fluctuations drop out from the action in the quadratic order. To avoid this problem we actually expand near the symmetry transform of the solution (\ref{sl2}), for which $\zeta_2$ and $\zeta_3$ are arbitrary parameters, not equal to $\pi/2$. Of course, the frequencies do not depend on them.
\begin{table}[h]
\beq \nn
\bea{c|c|c|c|c}
\toprule
            &{\rm\bf ws_1}             & {\rm\bf ws_2}       & {\rm\bf ac} & {\rm\bf multiplicity}\\
\midrule
sl(2)      & \bea{c} \omega^{S}_{n}\\\omega^{S}_{2n}/2 \eea & \bea{c} \omega^{S}_{n}\\\omega^{S}_{2(n+m/2)}/2 \eea & \bea{c} \omega^{S}_{n}\\ \omega^{S}_{2(n+m/2)}/2 \eea & \bea{c} \times 1 \\ \times 4\eea\\
\bottomrule
\eea
\eeq
\caption{\la{tabbos} \small Bosonic $\CP$ modes for the $sl(2)$ circular string in $AdS_4\times\CP^3$. The frequencies in the columns $\rm{ws}_1$ and $\rm{ws}_2$ correspond to the solution (\ref{sl1})\cite{McLRTs} and (\ref{sl2}), respectively. The column ${\rm ac}$ presents the result of the algebraic curve computation \cite{NG}.}
\end{table}

\enddocument